\documentclass[11pt]{article}

\usepackage{titlesec}
\titleformat*{\section}{\LARGE\bfseries}
\titleformat*{\subsection}{\Large\bfseries}
\titleformat*{\subsubsection}{\large\bfseries}

\makeatletter

\@addtoreset{equation}{section}

\@addtoreset{figure}{section}

\@addtoreset{table}{section}

\makeatother

\usepackage[numbers]{natbib}
\bibpunct[:]{[}{]}{,}{a}{}{,}

\usepackage{amsmath, amsthm,amsfonts, amssymb, bm, fancybox, multirow, hhline, psfrag, mathtools, color,algorithmic,algorithm}
\usepackage{graphicx}
\usepackage{here}
\usepackage{enumerate}
\usepackage{ascmac}
\usepackage{url}
\usepackage{color}
\theoremstyle{definition}

\newtheorem{remark}{Remark}[section]

\newcommand{\argmin}{\mathop{\rm arg~min}\limits}

\makeatletter
\renewcommand{\ALG@name}{アルゴリズム}
\makeatother
\allowdisplaybreaks[1]

\title{ 
	\huge{Improving prediction accuracy by choosing resampling distribution via cross-validation}\\[0.5cm]
}

\author{      
	\LARGE{Wataru Yoshida and Kei Hirose }\\[0.5cm]
	\LARGE{Kyushu University}\\[0.5cm]
}
\date{
	\LARGE{}
}

\begin{document}

\maketitle
\thispagestyle{empty}

\setcounter{page}{1}
\pagestyle{plain}

%%%%%%%%%%%%%%%%%%%%%%%%%%%%%%%%%%%%%%%%%%%%%%%%%%%%%%%%%%%%%%%%%%%%
\begin{abstract}
In a regression model, prediction is typically performed after model selection. The large variability in the model selection makes the prediction unstable. Thus, it is essential to reduce the variability in model selection and improve prediction accuracy. To achieve this goal, a parametric bootstrap smoothing can be applied. In this method, model selection is performed for each resampling from a parametric distribution, and these models are then averaged such that the distribution of the selected models is considered. Here, the prediction accuracy is highly dependent on the choice of a distribution for resampling. In particular, an experimental study shows that the choice of error variance significantly changes the distribution of the selected model and thus plays a key role in improving the prediction accuracy. We also observed that the true error variance does not always provide optimal prediction accuracy. Therefore, it would not always be appropriate to use unbiased estimators of the true parameters or standard estimators of the parameters for the resampling distribution. In this study, we propose employing cross validation to choose a suitable resampling distribution rather than unbiased estimators of parameters. Our proposed method was applied to electricity demand data. The results indicate that the proposed method provides a better prediction accuracy than the existing method.
\end{abstract}

\section{Introduction}
In regression modeling, estimation and prediction are typically performed after model selection. Reducing the variability in model selection is essential for achieving high accuracy. In addition, this variability should be considered when creating confidence or prediction intervals. The problem of appropriately handling the uncertainty caused by the model selection process is called selective inference \citep{taylor2015statistical}. For example, the selective inference in lasso estimation is researched by \citet{lee2016exact}.

To reduce variability in model selection and improve prediction accuracy, we may use parametric bootstrap smoothing. Parametric bootstrap smoothing performs model selection for each resampling from a parametric distribution, and these models are averaged such that the distribution of the selected models is considered. Thus, this method is expected to reduce the variability in model selection and improve prediction accuracy. A similar smoothing method using nonparametric bootstrapping is well-known in the field of machine learning under the name of bagging \citep{breiman1996bagging}. For these smoothing methods, \citet{efron2014estimation} proposed a formula for computing the standard errors of the estimators. Using this formula, we can construct the prediction interval in parametric bootstrap smoothing.

Suppose a sample follows the regression model $N(X\bm{\beta} , \sigma^{2} I_{n})$. To apply parametric bootstrap smoothing to estimate $\bm{\beta}$, the distribution for resampling must be set. A general approach is to choose $N(X\hat{\bm{\beta}}_{OLS} , \hat{\sigma}_{UB}^{2})$ as the resampling distribution \citep{mackinnon2006bootstrap}. Here, $\hat{\bm{\beta}}_{OLS}$ and $\hat{\sigma}_{UB}^{2}$ denote the ordinary least squares (OLS) estimator of the coefficient vector and the unbiased estimator of the variance based on $\hat{\bm{\beta}}_{OLS}$, respectively. However, in this study, we found that some distributions significantly outperform those based on OLS in terms of predictive accuracy. In fact, our simulation shows that the choice of error variance changes the distribution of the selected model and thus significantly affects prediction accuracy. We also discovered that choosing the true $\sigma^{2}$ for the resampling distribution did not always provide the optimal prediction accuracy. This suggests that it is inappropriate to use $\hat{\sigma}_{UB}^{2}$ for the resampling distribution, even though $\hat{\sigma}_{UB}^{2}$ is an unbiased estimator of the true variance. In addition, our simulation shows that it may not be optimal to set the mean vector to $X\hat{\bm{\beta}}_{OLS}$, especially when the true model is not full model. Therefore, we propose a method for choosing the resampling distribution by cross-validation, rather than model estimation. The proposed method searches for a distribution that specializes in improving the prediction of parametric bootstrap smoothing, whereas the existing methods select a distribution that fits the data. The proposed method is applied to predict the electricity demand. The results indicate that the proposed method provides better prediction accuracy than the existing methods.

The remainder of this paper is organized as follows. In Section \ref{sec:PBS}, we introduce the prediction and the prediction interval construction using parametric bootstrap smoothing. In Section \ref{sec:sigma_com}, we show the effect of the choice of the resampling distribution on the prediction accuracy from the experimental results of applying it to electricity demand data. Then, we propose a method for choosing a resampling distribution by cross-validation and show the results of an actual application to electricity demand data. In Section 4, we discuss through simulation why some choices of distribution provide better prediction accuracy than OLS-based distribution. In the simulation, the variance and mean vector of a resampling distribution were varied, and the prediction error and distribution of the selected models were examined. Finally, the conclusions are presented in Section 5.
%%%%%%%%%%%%%%%%%%%%%%%%%%%%%%%%%%%%%%%%%%%%%%%%%%%%%%%%%%%%%%%%%%%%
\section{Prediction using parametric bootstrap smoothing}\label{sec:PBS}
\subsection{Parametric bootstrap smoothing}\label{sec:PBS1}
Let ${\bf y}=(y_{1},\dots,y_{n})^T$ be the vector of $n$ observations. We assume that ${\bf y}$ follows the linear regression model:
\begin{align}\label{eq:reg_mod}
&{\bf y} = X\bm{\beta}+\bm{\varepsilon}, ~~~ \bm{\varepsilon} \sim N\left( 0,\sigma^2 I_{n} \right),
\end{align}
where $X$ is an $n \times p$ design matrix with $X:=({\bf x}_{1},\dots ,{\bf x}_{n})^{T}$ and $\bm{\beta}$ is a $p\times 1$ regression coefficient vector. Now, assume that we obtain ${\bf x}_{new}$, and $y_{new}$ follows the linear regression model:
\begin{align}
&y_{new} = {\bf x}_{new}^{T}\bm{\beta} + \varepsilon_{new}, ~~~ \varepsilon_{new} \sim N\left( 0,\sigma^2 \right). \nonumber
\end{align}
Our aim is to obtain the prediction value and prediction interval of $y_{new}$. We derive $\hat{\bm{\beta}}$, the estimator of $\bm{\beta}$ and predict $y_{new}$ as $\hat{\mu}( {\bf y} ) := {\bf x}_{new}^{T} \hat{\bm{\beta}}$. Model selection is conducted to derive $\hat{\bm{\beta}}$. For example, cross-validation \citep{stone1974cross}, AIC \citep{akaike1974new}, and sparse estimation can be considered. If $n$ is not sufficiently large for $p$, the model selection is unstable, and the prediction value of $y_{new}$ is also unstable. To address this problem, we use parametric bootstrap smoothing (see, for example, \citep{efron2014estimation}):
\begin{enumerate}
\item Compute $\hat{\bm{\beta}}_{OLS}=(X^TX)^{-1}X^T{\bf y}$ and $\hat{\sigma}^{2}$ that is an estimate of $\sigma^{2}$.
\item Generate bootstrap sample ${\bf y}_{1}^{*},\dots , {\bf y}_{B}^{*}$, where ${\bf y}_{i}^{*} \sim N(X\hat{\bm{\beta}}_{OLS} , \hat{\sigma}^{2} I_{n})$ for $i=1,\dots ,B$.
\item Derive $\hat{\bm{\beta}}({\bf y}_{1}^{*})$,\dots, $\hat{\bm{\beta}}({\bf y}_{B}^{*})$ that are estimates of $\bm{\beta}$ from each bootstrap sample ${\bf y}_{i}^{*}$. 
\item Compute the mean of estimates as $\hat{\bm{\beta}}_{PBS}( {\bf y} ):= \frac{1}{B}\sum_{b=1}^{B}\hat{\bm{\beta}}({\bf y}_{b}^{*})$, and we get the prediction of $y_{new}$ as follows:
\begin{align}\label{eq:PBSforecast}
&\hat{\mu}_{PBS}({\bf y}) := x_{new}^{T}\hat{\bm{\beta}}_{PBS}( {\bf y} ).
\end{align}
\end{enumerate}
In Step 3, note that the result of the model selection varies with each bootstrap sampling, that is, even if $\hat{{\beta}}_{k}({\bf y}_{i}^{*})=0$, it may be $\hat{{\beta}}_{k}({\bf y}_{j}^{*})\neq 0$ ($i\neq j$). Here, $\hat{{\beta}}_{l}({\bf y}_{m}^{*})$ denotes $l$th component of $\hat{\bm{\beta}}({\bf y}_{m}^{*})$. In Step 4, the randomness of model selection is smoothed.

\subsection{Prediction interval}\label{sec:PBSCI}
We consider a $100(1-\alpha )$ \% prediction interval for $y_{new}$ similar to the confidence interval derivation process proposed by \citet{efron2014estimation}. First, the estimator of $V(\hat{\mu}_{PBS}({\bf y}) )$ is derived by using the delta method as follows:
\begin{align}\label{eq:VhatPBS}
&\hat{V}(\hat{\mu}_{PBS}({\bf y}) ) := \frac{1}{\hat{ \sigma }^{2}} \widehat{Cov}^{T} (X^{T}X)^{-1} \widehat{Cov},
\end{align}
where
\begin{align*}
&\widehat{Cov}:=\frac{1}{B}\sum_{b=1}^{B}(\hat{\mu}( {\bf y}_{b}^{*} ) - \hat{\mu}_{PBS}({\bf y}) )(X^{T}{\bf y}_{b}^{*} - X^{T}\bar{{\bf y}}^{*}), ~~~ \bar{{\bf y}}^{*} := \frac{1}{B}\sum_{b=1}^{B}{\bf y}_{b}^{*},\ \hat{\mu}( {\bf y}_{b}^{*}):=x_{new}^{T} \hat{\bm{\beta}}({\bf y}_{b}^{*}).
\end{align*}
Next, $V(\varepsilon_{new})$ is estimated as $\frac{1}{n-p} \| {\bf y} - X\hat{\bm{\beta}}_{PBS}({\bf y}) \|_{2}^{2} $. Using the above estimators, the $100(1-\alpha )$ \% prediction interval of $y_{new}$ is derived as follows:
\begin{align}\label{eq:PBSCI}
&\hat{\mu}_{PBS}({\bf y}) \pm z_{ \alpha / 2}\sqrt{ \hat{V}(\hat{\mu}_{PBS}({\bf y}) ) + \frac{1}{n-p} \| {\bf y} - X\hat{\bm{\beta}}_{PBS}({\bf y}) \|_{2}^{2} },
\end{align}
where $z_{\alpha/2}$ is the $100(1- \frac{\alpha}{2} )$th percentile of the standard normal distribution.
%%%%%%%%%%%%%%%%%%%%%%%%%%%%%%%%%%%%%%%%%%%%%%%%%%%%%%%%%%%%%%%%%%%%
\section{Effect of resampling distribution parameters on prediction accuracy}\label{sec:sigma_com}
\subsection{Application of parametric bootstrap smoothing to electricity demand data}\label{sec:denki}
We applied parametric bootstrap smoothing to the Tokyo electricity demand data and verified the validity of the prediction accuracy. The source of this data is \url{https://www.tepco.co.jp/forecast/html/download-j.html}.

\begin{figure}[H]
\centering
\includegraphics[width=13cm, bb=0 0 842 572]{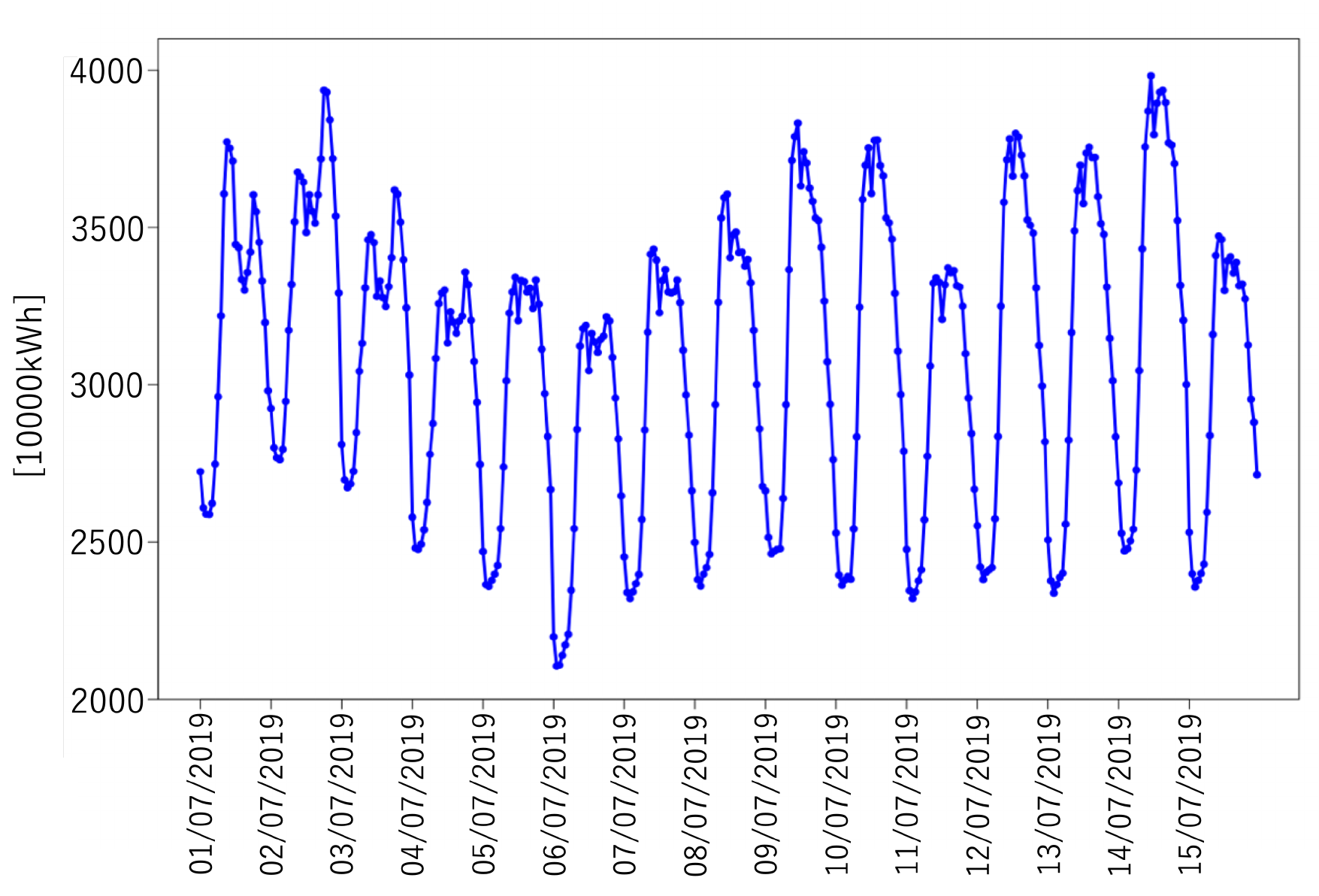}
\caption{Electricity demand data (hourly)}
\label{zu1_bootstrap}
\end{figure}

Let $y_{ij}$ be electricity demand at $j:00$ on day $i$ ($j=1,\dots ,24$). We use a regression model with the temperature information proposed by \citet{hirose2021interpretable}. The source of the temperature information is \url{https://www.data.jma.go.jp/gmd/risk/obsdl/index.php}. Specifically, the following model was considered:
\begin{align}\label{eq:denryoku_mod}
&y_{ij} = \sum_{t=1}^{T}\alpha_{jt}y_{(i-t)j} + \sum_{m=1}^{M}\sum_{q=1}^{Q}\gamma_{qm}h_{q}(j)g_{m}(s_{i}) + \varepsilon_{ij} ~~~~~ \varepsilon_{ij}\sim N(0,\sigma^{2}),
\end{align}
where $s_{i}$ is the average temperature on day $i$; $h_{q}(j)$ is the cyclic B-spline basis function; and $g_{m}(s_{i})$ is the B-spline basis function. The estimated regression coefficients are $\alpha_{jt}$ and $\gamma_{qm}$ (hereafter, the regression coefficients are collectively written as $\bm{\beta}$). The following experiments are conducted using this model.

\begin{itemize}
\item[] [Experiment overview]
\item Four candidate models are prepared by changing the number of B-spline bases. The dimensions $p$ of $\bm{\beta}$ in each model are 121, 146, 146, 196, respectively.
\item The estimate $\hat{\bm{\beta}}$ is computed by ridge regression \citep{hoerl1970ridge} to make a prediction. The model and regularization parameter $\lambda$ are selected by generalized cross-validation, respectively. The prediction interval is derived by regarding $V(\hat{\bm{\beta}}) = \sigma^{2} (X^{T}X + \lambda I_{p})^{-1} X^{T}X (X^{T}X + \lambda I_{p})^{-1}$, without considering the randomness of the model selection.
\item The parametric bootstrap smoothing is applied to the above ridge regression. Here, bootstrap sample size $B$ is 500, and $\hat{\sigma}^{2}$, which is in step 1 of the parametric bootstrap smoothing procedure introduced in Section \ref{sec:PBS1}, is determined as $\hat{\sigma}_{UB}^{2} := \frac{1}{n-p} \| {\bf y} - X\hat{\bm{\beta}}_{OLS} \|_{2}^{2}$.
\item The prediction values and the prediction intervals of $y_{i1},\dots ,y_{i24}$ are computed for each day $i=1,\dots ,365$ (1/4/2019$\sim$31/3/2020). The training data used in this prediction is the last 15 days of the data which are on the same day of the week as the prediction date $i$. Then, we compared the prediction accuracy between ridge regression and parametric bootstrap smoothing.
\end{itemize}
Table \ref{table1} compares the prediction and prediction intervals for accuracy between ridge regression and parametric bootstrap smoothing. The results show that although the prediction accuracy was improved, the prediction interval did not improve, but rather worsened. However, we conducted further experiments and found that the accuracy varied significantly depending on the choice of $\hat{\sigma}^{2}$ used for resampling in the parametric bootstrap smoothing.

\begin{table}[htb]
\centering
  \caption{Comparison of two methods}
  \label{table1}
  \begin{tabular}{|l||c|c|c|c|c|c|c|}  \hline
    {} & Ridge regression & Parametric bootstrap smoothing \\ \hline \hline
    Mean-squared prediction error & 147051.5 & 126762.8 \\ \hline
    Percentage of observed values included & \multirow{2}{*}{72.8\%} & \multirow{2}{*}{70.7\%} \\
    in prediction interval & & \\ \hline
    \end{tabular}
\end{table}

\subsection{Experiments comparing the prediction accuracy for each different $\hat{\sigma}^2$} \label{sec:sim denki}
\begin{figure}[H]
\centering
\includegraphics[width=13cm, bb=0 0 843 571]{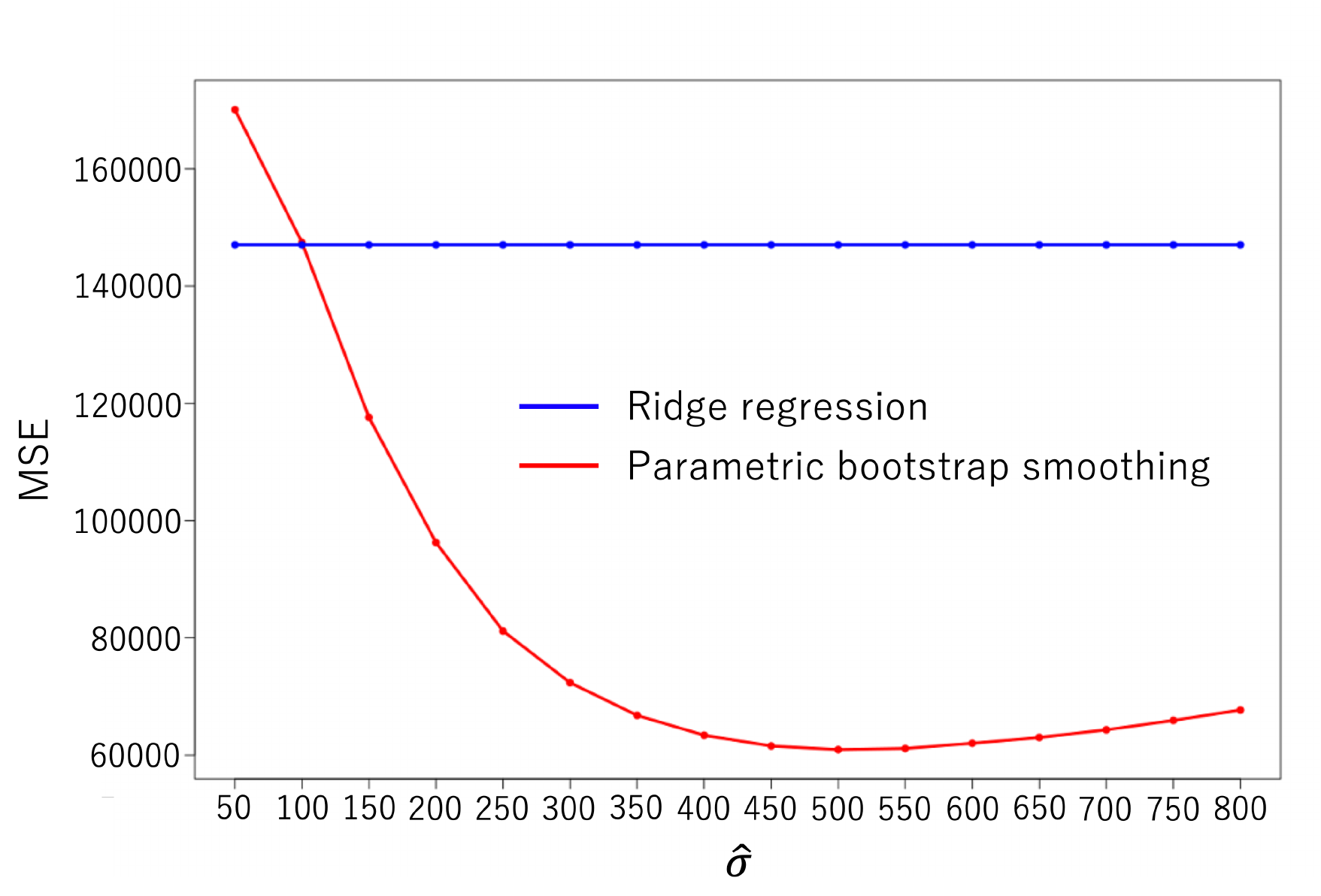}
\caption{Mean-squared prediction error}
\label{zu2_bootstrap}
\end{figure}

\begin{figure}[H]
\centering
\includegraphics[width=13cm, bb=0 0 833 594]{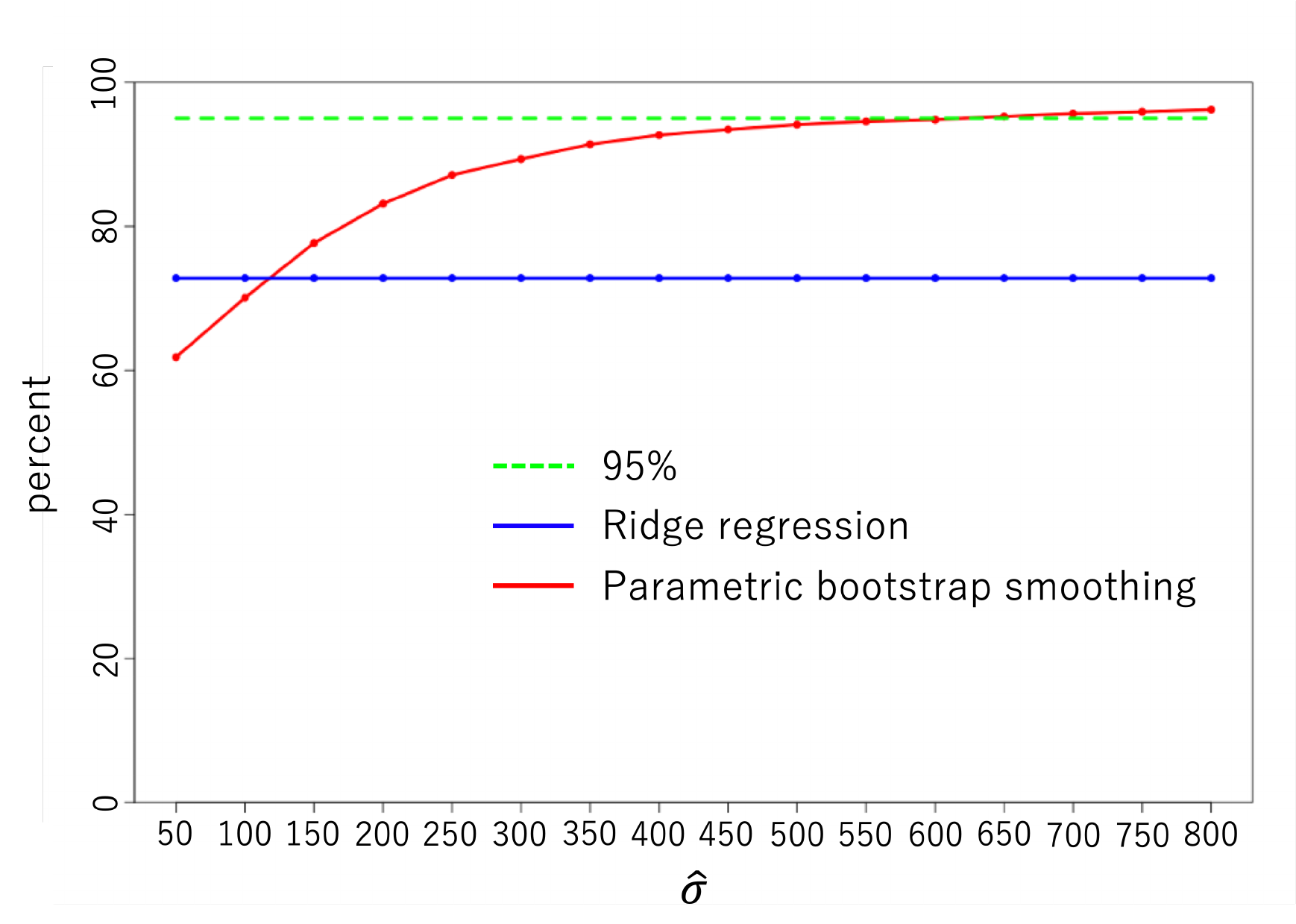}
\caption{Percentage of observed values included in prediction interval}
\label{zu3_bootstrap}
\end{figure}

In Section \ref{sec:denki}, we chose $\hat{\sigma}^{2}$ as $\hat{\sigma}_{UB}^{2} =\frac{1}{n-p} \| {\bf y} - X\hat{\bm{\beta}}_{OLS} \|_{2}^{2}$; but in this Section, we give $\hat{\sigma}^{2}$ in advance and conduct the same experiment as in Section \ref{sec:denki} to test the affect of the choice of $\hat{\sigma}^2$. The choice of $\hat{\sigma}^2$ varies as $\hat{\sigma}^{2}=50^{2}, 100^{2}, \dots ,800^{2}$. Figure \ref{zu2_bootstrap} and \ref{zu3_bootstrap} show the mean squared prediction error and percentage of observed values included in the prediction interval, respectively. The results show that the choice of $\hat{\sigma}^{2}$ has a significant effect on prediction accuracy. In particular, for some variance values, the prediction accuracy was significantly improved over choosing the variance $\hat{\sigma}_{UB}^{2}$.

\subsection{Choosing of resampling distribution by cross-validation}\label{sec:muCV}
We have evaluated the effect of the choice of $\hat{\sigma}^{2}$. Now, we also focus on the mean vector. Specifically, we consider resampling from the following distribution: 
\begin{align}\label{bunpu_PBS}
&N \Bigl( \hat{\gamma} X\hat{\bm{\beta}}_{OLS} + (1-\hat{\gamma}) {\bf y} , \hat{\sigma}^{2} I_{n} \Bigl), \text{ where } 0 \leq \hat{\gamma} \leq 1.
\end{align}
The properties of this mean vector are discussed in Section \ref{sec:sim}.

\begin{figure}[H]
\centering
\includegraphics[width=13cm, bb=0 0 842 576]{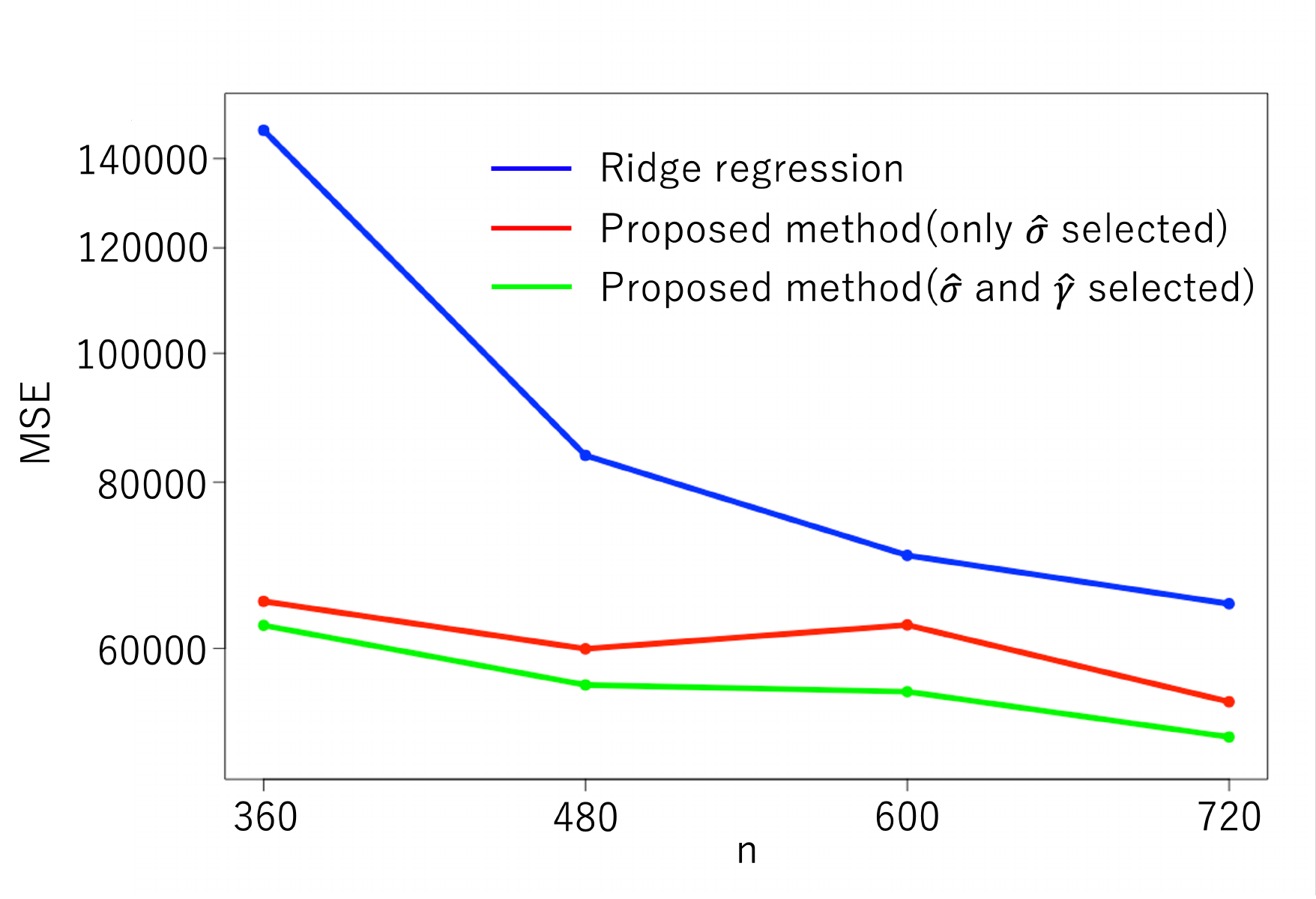}
\caption{Mean-squared prediction error}
\label{zu7_bootstrap}
\end{figure}

\begin{figure}[H]
\centering
\includegraphics[width=13cm, bb=0 0 809 593]{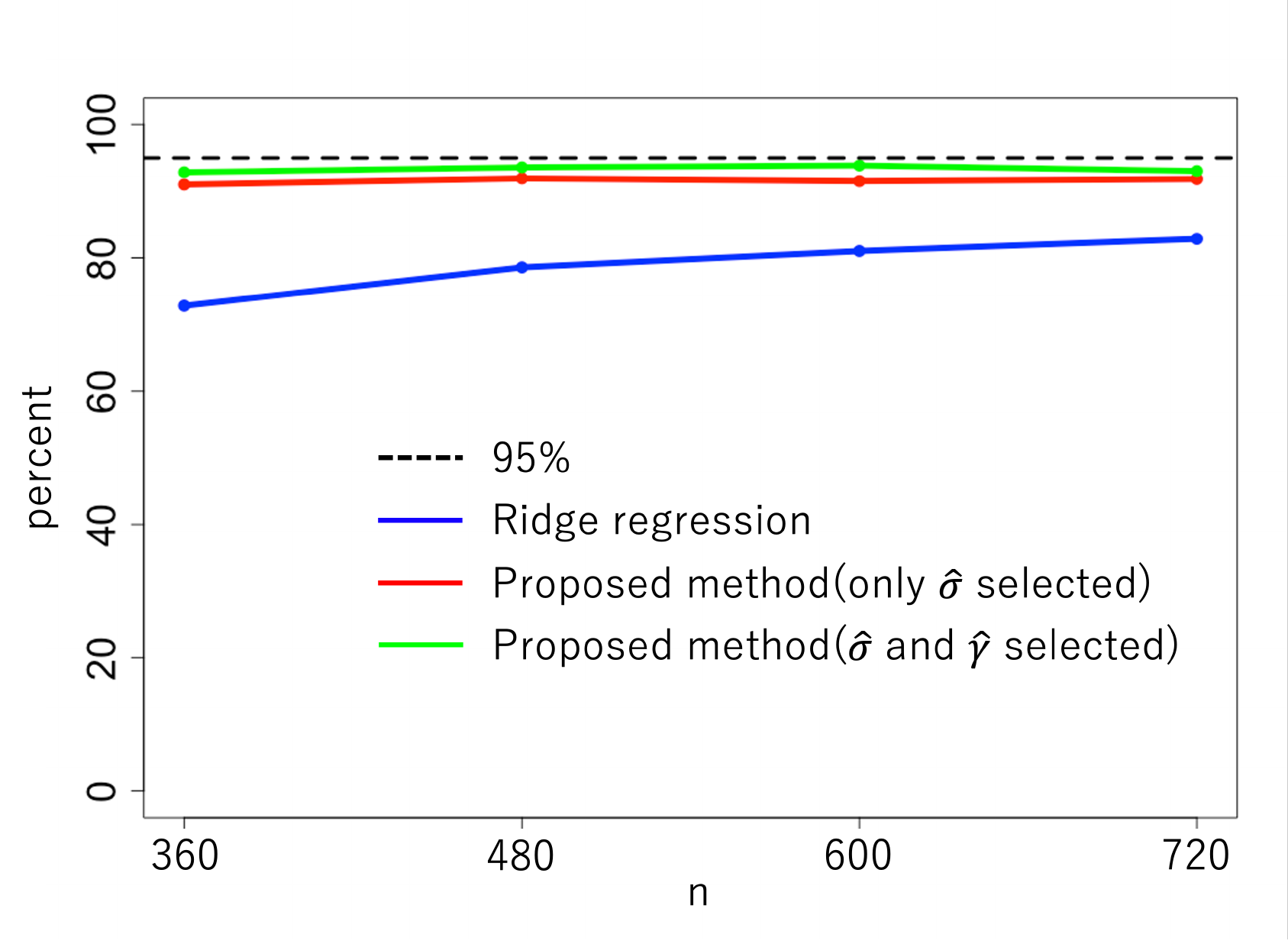}
\caption{Percentage of observed values included in prediction interval}
\label{zu6_bootstrap}
\end{figure}

We aim to find $\hat{\sigma}^{2}$ and $\hat{\gamma}$, and choose a resampling distribution with better accuracy than the distribution based on OLS estimation. Here, we propose a method for choosing $\hat{\sigma}^{2}$ and $\hat{\gamma}$ using cross validation, as follows:
\begin{enumerate}
\item Split ${\bf y}$ into $K$ pairs, and let ${\bf y}=({\bf y}_{(1)}^T,\ ...,\ {\bf y}_{(K)}^T)^T$.
\item Give $\hat{\sigma}^2$ and $\hat{\gamma}$ candidates as $\bar{\sigma}^2_{1},\ ...,\  \bar{\sigma}^2_{T}$ and $\bar{\gamma}_{1},\ ...,\  \bar{\gamma}_{S}$.
\item The parametric bootstrap smoothing with ${\bf y}^{(k)}$ is conducted to compute $\hat{\mu}_{PBS}({\bf y}^{(k)}, \bar{\sigma}^2_{i}, \bar{\gamma}_{j})$, which is the prediction of ${\bf y}_{(k)}$, for each $\bar{\sigma}^2_{i}$, $\bar{\gamma}_{j}$, and $k$. Here, ${\bf y}^{(k)}$ is ${\bf y}$ with ${\bf y}_{(k)}$ removed. Bootstrap samples are generated from $N \Bigl( \bar{\gamma}_{j} X^{(k)} \hat{\bm{\beta}}_{OLS} + (1-\bar{\gamma}_{j}) {\bf y}^{(k)} , \bar{\sigma}^2_{i} I_{n^{(k)} } \Bigl) $, where $X^{(k)}$ denotes the submatrix of $X$ corresponding to ${\bf y}^{(k)}$, and $n^{(k)}$ denotes the dimension of ${\bf y}^{(k)}$.
\item Choose $\hat{\sigma}^2$ and $\hat{\gamma}$ as follows:
\begin{align*} (\hat{\sigma}^2,\ \hat{\gamma}) = \argmin_{\bar{\sigma}^2_{i},\ \bar{\gamma}_{j}} {\sum_{k=1}^{K} {\|{\bf y}_{(k)} - \hat{\mu}_{PBS}({\bf y}^{(k)}, \bar{\sigma}^2_{i} , \bar{\gamma}_{j})\|^2_2}},
\end{align*}
and the parametric bootstrap smoothing using chosen $\hat{\sigma}^2$ and $\hat{\gamma}$ is conducted.
\end{enumerate}
When computing the prediction interval, ${V}(\hat{\mu}_{PBS}({\bf y}) )$ is estimated as
\begin{align}\label{vhat_2}
&\hat{V}(\hat{\mu}_{PBS}({\bf y}) ) := \frac{1}{\hat{ \sigma }^{2}} \widehat{Cov}^{T} \{ \hat{\gamma} X(X^{T}X)^{-1}X^{T} + (1 - \hat{\gamma})I_{n} \}^{2} \widehat{Cov},
\end{align}
where $\widehat{Cov}:=\frac{1}{B}\sum_{b=1}^{B}(\hat{\mu}( {\bf y}_{b}^{*} ) - \hat{\mu}_{PBS}({\bf y}) )({\bf y}_{b}^{*} - \bar{{\bf y}}^{*})$. As in \eqref{eq:VhatPBS}, we derive this estimator using the delta method. We conducted the same experiment as that in Section \ref{sec:denki} using this method. The number of candidates for $\hat{\sigma}^2$ and $\hat{\gamma}$ is 50 and 6, respectively, and the bootstrap sample size is 500. The data used for the prediction were changed over 15, 20, 25, and 30 days (sample sizes $n = 360,480,600,720$). Figures \ref{zu7_bootstrap} and \ref{zu6_bootstrap} show the mean squared prediction error and percentage of observed values included in the predicted interval for each sample size, respectively. In addition to the ridge regression and the proposed method, a method that selects only $\hat{\sigma}^2$ by cross-validation was also compared (i.e., fixed $\hat{\gamma}=1$). The results showed that the proposed method gave the best results, independent of the sample size. It was found to be particularly effective compared with ridge regression when the sample size was small.

%%%%%%%%%%%%%%%%%%%%%%%%%%%%%%%%%%%%%%%%%%%%%%%%%%%%%%%%%%%%%%%%%%%%
\section{Simulation study}\label{sec:sim}
\begin{figure}[H]
\centering
\includegraphics[width=18cm, bb=0 0 1123 640]{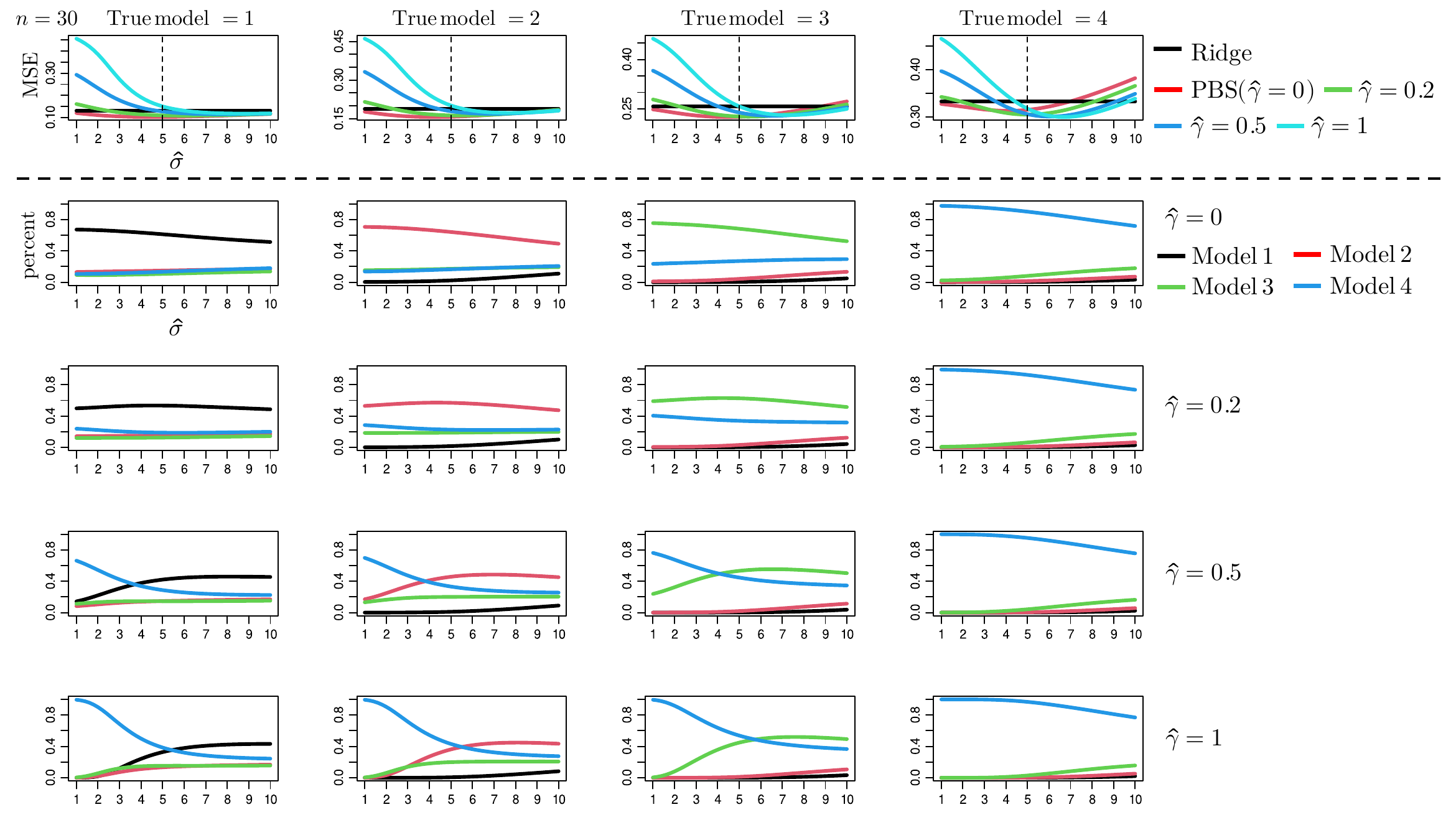}
\caption{Mean-squared error of estimation and percentage of each model selected in $n=30$}
\label{zu11_1}
\end{figure}

\begin{figure}[H]
\centering
\includegraphics[width=18cm, bb=0 0 1123 640]{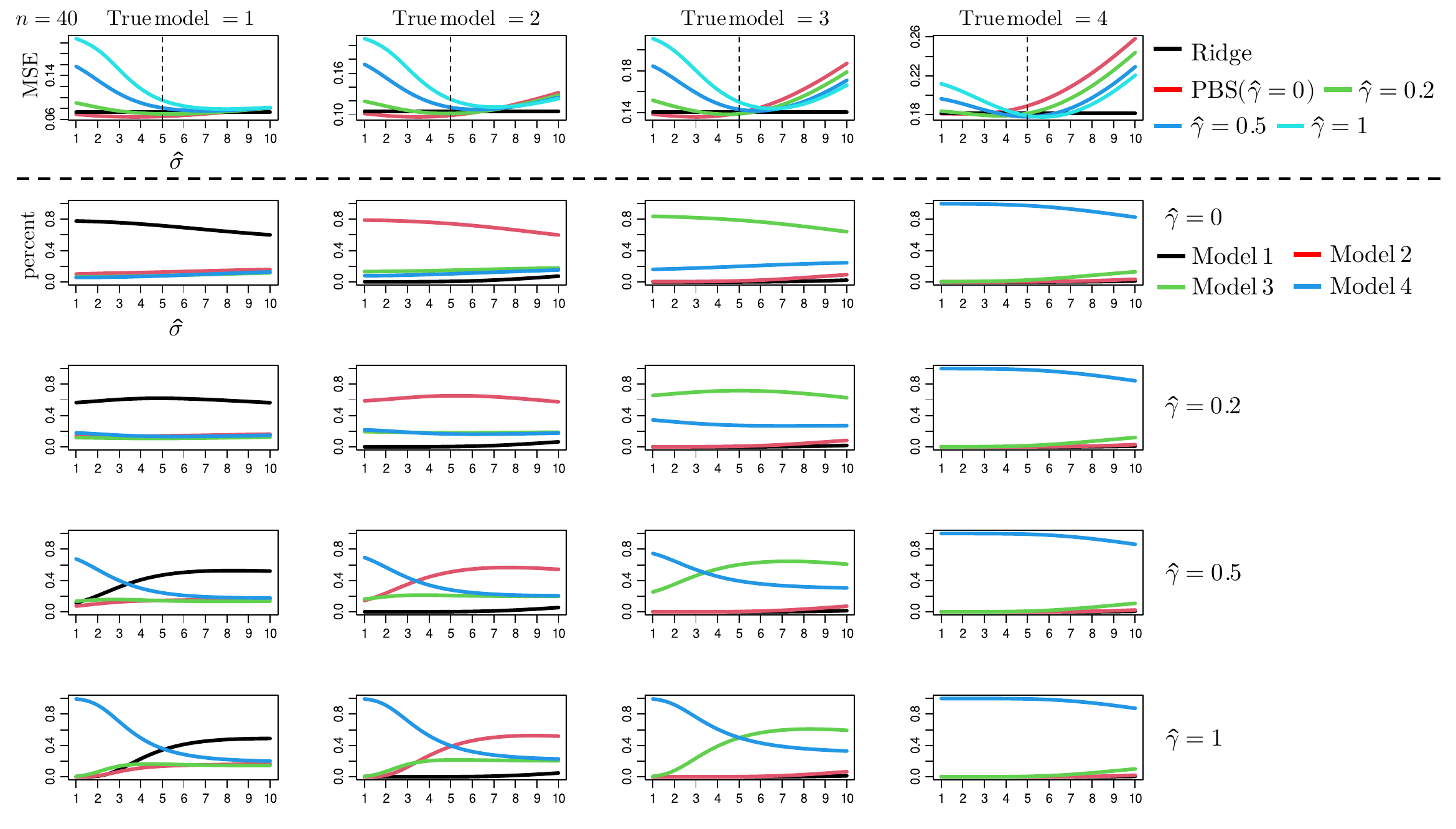}
\caption{Mean-squared error of estimation and percentage of each model selected in $n=40$}
\label{zu11_2}
\end{figure}

In this Section, we discuss, through simulation, why some choices of $\hat{\sigma}^2$ provide better prediction accuracy than $\hat{\sigma}_{UB}^{2}$. We also consider changing the mean vector and its effect on prediction accuracy. First, random numbers were independently generated from a uniform distribution $U(-5,5)$ to create an $n \times 20$ design matrix $X$. Subsequently, we considered four models. The $j$th model is as follows:
\begin{align*}
&{\bf y} = {\bf 1}_{n} + X\bm{\beta}_{j}+\bm{\varepsilon}, ~~~ \bm{\varepsilon} \sim N\left( 0,5^2 I_{n} \right),
\end{align*}
where $\bm{\beta}_{j} = ( {\bf 1}_{5j}^{T} , {\bf 0}_{20-5j}^{T} )^{T}$ for $j = 1,2,3,4$, and ${\bf 1}_{m}$, ${\bf 0}_{m}$ denote $m \times 1$ vectors with ${\bf 1}_{m}=(1, \dots ,1)^{T}$, ${\bf 0}_{m}=(0, \dots ,0)^{T}$. In the simulation, {\bf y} was generated from one of the four different models and a similar experiment to that in Section \ref{sec:sim denki}. 
Here, we consider resampling from \eqref{bunpu_PBS} and varying the mean vector in addition to $\hat{\sigma}^{2}$.
\begin{remark}
Let ${\bf y}^{*}=\hat{\gamma} X\hat{\bm{\beta}}_{OLS} + (1-\hat{\gamma}) {\bf y} + \bm{\varepsilon}^{*}$ be the bootstrap sample generated from \eqref{bunpu_PBS}. The ridge estimator based on model $j$ can then be computed as follows:
\begin{align*}
&(X_{j}^{T} X_{j} + \lambda I_{5j} )^{-1} X_{j}^{T} {\bf y}^{*} = (X_{j}^{T} X_{j} + \lambda I_{5j} )^{-1} X_{j}^{T} (\hat{\gamma} X\hat{\bm{\beta}}_{OLS} + (1-\hat{\gamma}) {\bf y} + \bm{\varepsilon}^{*}),
\end{align*}
where $X_{j}$ denotes $n \times 5j$ design matrix in model $j$ with $X_{j} := X (I_{5j} , 0)^{T}$. Here, since
\begin{align*}
&(X_{j}^{T} X_{j} + \lambda I_{5j} )^{-1} X_{j}^{T} X\hat{\bm{\beta}}_{OLS}
\\
& = (X_{j}^{T} X_{j} + \lambda I_{5j} )^{-1}  (I_{5j},0) X^{T} X (X^{T}X)^{-1} X^{T} {\bf y} = (X_{j}^{T} X_{j} + \lambda I_{5j} )^{-1} X_{j}^{T} {\bf y},
\end{align*}
it holds that $(X_{j}^{T} X_{j} + \lambda I_{5j} )^{-1} X_{j}^{T} {\bf y}^{*} = (X_{j}^{T} X_{j} + \lambda I_{5j} )^{-1} X_{j}^{T} ({\bf y} + \bm{\varepsilon}^{*})$. That is, $\hat{\gamma}$ does not affect the ridge estimation itself for any bootstrap sample. Therefore, the twice-shrinking noted in Section 7A of \citep{efron2014estimation} can be avoided. For each bootstrap sample, $\hat{\gamma}$ is only related to the results of model and $\lambda$ selection by cross-validation. In particular, when $\hat{\gamma}$ is large, the mean vector is close to $X\hat{\bm{\beta}}_{OLS}$. Therefore, the full model is more likely to be selected for cross-validation. Thus, $\hat{\gamma}$ would be interpreted as a measure of the trust of the prepared full model.
\end{remark}
The choices of $\hat{\sigma}^2$ and $\hat{\gamma}$ varied as $\hat{\sigma}^2 = 1^{2}, 1.2^{2}, 1.4^{2} , \dots , 10^{2}$ and $\hat{\gamma} = 0,0.2,0.5,1$, respectively. This simulation was repeated 1000 times to evaluate the accuracy of the estimation of $\bm{\beta}_{j}$ and the constant term. Figures \ref{zu11_1} and \ref{zu11_2} show the results for $n=30$ and $40$, respectively (the results for $n>40$ are similar to those of $n=40$). Column $j$ of the figure shows the results when the true model is $j$. The top figure compares the mean squared estimation errors, and the four figures below show the percentage of each model selected in parametric bootstrap smoothing for each $\hat{\gamma}$. The results of MSE show that the optimal $\hat{\sigma}^2$ depends on $n$, $\hat{\gamma}$, and the true model. In addition, compared to ridge estimation, parametric bootstrap smoothing is more effective when $n$ is small. We find that the mixing ratio of the model selection varies significantly, depending on $\hat{\sigma}^2$ and $\hat{\gamma}$. In particular, $\hat{\gamma}$ has a significant influence on the probability of selecting the full model. However, when $\hat{\sigma}^2$ is large, the mixing ratio is similar, regardless of $\hat{\gamma}$. Note that a high probability of selecting the true model does not necessarily lead to optimal prediction accuracy. For example, looking at the results for the case in which the true model is Model 4, it seems that a smaller $\hat{\sigma}^2$ is optimal in terms of the probability that Model 4 is selected, but not in terms of MSE. This fact would be because $\hat{\sigma}^2$ and $\hat{\gamma}$ affect not only the ratio of model selection but also the ratio of the selection of the regularization parameter $\lambda$. For example, considering the case $\hat{\sigma}^2 = 0$ and $\hat{\gamma} = 1$, the bootstrap sample is always $X\hat{\bm{\beta}}_{OLS}$, thus the full model is always selected, but the $\lambda$ is also always 0. Consequently, $\hat{\bm{\beta}}_{PBS} = \hat{\bm{\beta}}_{OLS}$ and the variance in $\hat{\bm{\beta}}_{PBS}$ increases.

As mentioned above, the optimal $\hat{\sigma}^2$ varies complexly by various factors, and it is not always the true variance, which may be the reason why some choices of $\hat{\sigma}^2$ provide better prediction accuracy than $\hat{\sigma}_{UB}^{2}$, which is an unbiased estimator of $\sigma^2$. Additionally, it is not always optimal to set the mean vector to $X\hat{\bm{\beta}}_{OLS}$ ($\hat{\gamma} = 1$). Overall, it may be better not to determine a resampling distribution based on model estimation.

%%%%%%%%%%%%%%%%%%%%%%%%%%%%%%%%%%%%%%%%%%%%%%%%%%%%%%%%%%%%%%%%%%%%
\section{Concluding remarks}
We performed a Monte Carlo simulation and electricity demand prediction to investigate the effectiveness of parametric bootstrap smoothing. The results show that the choice of variance $\hat{\sigma}^2$ when generating bootstrap samples has a significant effect on prediction accuracy. In particular, we found that true variance does not always provide optimal prediction accuracy. We focused on this and the mean vector, and then proposed a method to conduct parametric bootstrap smoothing with a resampling distribution chosen by cross-validation. In our simulation, the proposed method showed better prediction accuracy than the method using a resampling distribution based on OLS estimation.

In future studies, we will consider developing a faster method. The proposed method requires $K$ $\times$ $T$ $\times$ $S$ times parametric bootstrap smoothing, which is computationally expensive. Here, $K$ is the number of data splits in the cross-validation and $T$ and $S$ are the numbers of $\hat{\sigma}^{2}$ and $\hat{\gamma}$ candidates, respectively. We will also consider researching how the $\hat{\sigma}$ and mean vector selected by cross-validation are related to the true $\sigma$ and mean vector.
%%%%%%%%%%%%%%%%%%%%%%%%%%%%%%%%%%%%%%%%%%%%%%%%%%%%%%%%%%%%%%%%%%%%
\section*{Acknowledgments}
This work was supported by JSPS KAKENHI Grant Numbers JP22J20435 (W.Y.), JP23K11007, JP23K01333, JP23H04474, JP23H00466, and JP22H01139 (K.H.), WISE program (MEXT) at Kyushu University, and JST-Mirai Program Grant Number JPMJMI18A2.

\renewcommand{\refname}{References}

\bibliographystyle{plainnat}
\bibliography{reference_combine}

\end{document}